\documentclass[a4paper,11pt]{article}
\usepackage[utf8x]{inputenc}
\begin{document}

\centerline{\Large \bf Novel and Unique Expression for the} 
\centerline{\Large \bf Radiation Reaction Force,}
\centerline{\Large \bf Relevance of Newton's Third Law and Tunneling}

\begin{center}
{\bf Dieter Gromes and Eduard Thommes}\\[3 ex]Institut f\"ur
Theoretische Physik der Universit\"at Heidelberg\\ Philosophenweg 16,
D-69120 Heidelberg \\ E - mail: d.gromes@thphys.uni-heidelberg.de,
e.thommes@thphys.uni-heidelberg.de \\
 {\em March 2014}
\end{center} 
\vspace{0.4cm}

{\bf Abstract:} We derive the radiation reaction by taking into
account that the acceleration of the charge is caused by the interaction 
with some heavy source particle. In the non relativistic case this 
leads, in contrast to the usual approach, immediately to a result
which is Galilei invariant. Simple examples show that there can be small
regions of extremely low velocity where the energy requirements cannot
be fulfilled, and which the charged particle can only cross by quantum mechanical 
tunneling. We also give the relativistic generalization which appears unique.
The force is a four-vector, but only if the presence of the source is
taken into account as well. It contains no third derivatives of the 
position as the Lorentz-Abraham-Dirac equation, and consequently no 
run away solutions. All examples considered so far give reasonable
results.

\vspace{0.4cm}

\section{Introduction: What has gone wrong?}\
``The problem of radiation reaction and the self force is the oldest
unsolved mystery in physics'' as stated in the review article of 
Hammond \cite{Hammondrev}. We refer to this article for details 
about history, suggested solutions, and literature. 

Let us very briefly recall the standard 
derivation of the reaction force, the short comings in this derivation,
 as well as the problems arising 
with the result obtained in this way. 

The starting point is the well known Larmor formula
for the energy loss per time of a charged particle, accelerated
by some external force:
\begin{equation}
 \frac{dE}{dt} = - \frac{2}{3} \frac{e²}{c³} \dot{\bf v}².
\end{equation}
One argues that this energy loss has to be compensated by 
the power due to a radiation reaction force ${\bf f}_{rad}$ which acts
on the particle:
\begin{equation}
 {\bf f}_{rad} \, {\bf v} = - \frac{2}{3} \frac{e²}{c³} \dot{\bf v}².
\end{equation}
Subsequently one integrates this relation over some time interval and performs 
a partial integration:

\begin{equation}
 \int _{t_1}^{t_2} {\bf f}_{rad} \, {\bf v} dt 
= - \frac{2}{3} \frac{e²}{c³} \int _{t_1}^{t_2} \dot{\bf v}^2 dt
= \frac{2}{3} \frac{e²}{c³} \Big(\int _{t_1}^{t_2} \ddot{\bf v} {\bf v}dt
-   \dot{\bf v} {\bf v} |_{t_1}^{t_2}\Big).
\end{equation}
One now assumes that $\dot{\bf v} {\bf v}$ vanishes at the end points
and boldly concludes
\begin{equation}
 {\bf f}_{rad} = \frac{2}{3} \frac{e²}{c³} \ddot{\bf v}.
\end{equation}
Finally one can write down the relativistic generalization $F^\mu_{rad}$ of the 
radiation reaction force:

\begin{equation}
F^\mu_{rad} =  \frac{2}{3} \frac{e²}{c^3}
\big(\frac{d² u^\mu}{d\tau^2} + \frac{du_\nu}{d\tau} \frac{du^\nu}{d\tau}
\frac{u^\mu}{c^2}\big),
\end{equation}
with $u^\mu$ the velocity four vector and $\tau $ the proper time. 
The second term has been added in order to ensure $u_\mu F^\mu_{rad}=0.$ 
The corresponding equation of motion is often called the Lorentz-Abraham-Dirac 
(LAD-) equation \cite{Lorentz}\cite{Abraham}\cite{Dirac}.

Maliciously speaking one may say that the Larmor formula (1) is the very last
formula in this chain which is correct. Already (2), though presented
in an innumerable number of text books, must obviously be wrong  - from the simple
reason that it is not Galilei invariant! Under a 
non relativistic boost, i.e. a Galilei transformation of the form 
${\bf x} = {\bf x}' + {\bf w} t$, forces and accelerations are invariant, while 
velocities change according to ${\bf v} = {\bf v}' + {\bf w}$. Clearly
eq. (2) is not Galilei invariant. 

The assumptions
made after the partial integration are as well dubious and rarely ever fulfilled. 
The integrated terms $\dot{\bf v} {\bf v} |_{t_1}^{t_2}$ are again not
Galilei invariant and only vanish for rather
special cases. Nevertheless one appears astonished when the resulting equation, 
applied to situations where the assumption is definitely not fulfilled,
gives unphysical run away solutions. It is also not mandatory to conclude from 
the equality of the integrals to the equality of the integrands. And, finally, 
one only gets an information on the component of the force parallel to the velocity.
The resulting equation of motion is also strange because it contains
the third time derivative of the position, a rather unfamiliar feature. 
For the initial value problem one has to prescribe not only position and 
velocity, but also the acceleration.

In the present work we will approach and solve the problem in a direct way. 
An essential point is 
to give up the unphysical concept of an external field which is implicitly
present in the usual procedure. Instead we assume that 
the ``external'' force arises through the interaction with a second (very heavy) 
source particle. We work directly with the equations for momentum and energy,
without performing partial integrations or other manipulations, or needing any
assumptions about the internal structure of the charge. 

In the non relativistic case treated in sect. 2 our procedure corrects 
immediately the wrong equation (2), introduces the difference between the velocities
of particle and source, and leads to a result which is Galilei invariant.
No  third derivatives $\stackrel{...}{\bf x}$
appear, one has the standard initial value problem of prescribing position 
and momentum, no run away solutions show up. For applications we assume, of
course, that the mass of the source particle is very large, so it becomes
 essentially a static source.
We apply our equation to three simple problems: Constant external
force, harmonic oscillator, and circular motion.
Some of the solutions show a remarkable property, at first sight a serious defect, 
but indeed an inevitable phenomenon which can be well understood and will be 
explained in detail later. In some (very tiny)
regions there is no solution because the energy requirements cannot
be fulfilled. The mathematics reacts by producing a complex solution in
these regions. This indicates that we have to leave the field
of classical physics here and must enter the world of quantum mechanics.
The charged particle has to tunnel through a classically forbidden region.

In sect. 3 we generalize to relativistic motion. We construct a four-force
which we consider as unique. It has
all the properties which are required.
The manifestly covariant form, as was to be expected from the 
non relativistic case, can only be obtained if one considers the source particle
as well. 
We apply our equation to the linear relativistic motion in a constant 
electric field. This leads to a reasonable result.

Sect. 4 summarizes the properties of the radiation reaction force
obtained here.

\section{Non relativistic equation}\
We abandon the unphysical concept of an external force as reason for the acceleration 
and replace it by the force 
created by a large mass $M$ situated at ${\bf X}$, which interacts with the considered particle
 of mass $m$ and charge $e$ via a potential $\varphi ({\bf x} - {\bf X})$. The velocities of 
the particle and
the source are denoted by ${\bf v}$ and ${\bf V}$. The source particle $M$ has no charge for
simplicity.

For non relativistic motion one can ignore the radiation of momentum, there is 
only radiation of energy. Momentum conservation then gives  
${m\dot{\bf v} + M\dot{\bf V}} = 0$, while the energy loss 
per time becomes

\begin{eqnarray}
& & \frac{d}{dt} \Big(\frac{m}{2} v² + \frac{M}{2} V² + \varphi  \Big) \nonumber\\
& = & m {\bf v} \dot{\bf v} + M {\bf V} \dot{\bf V} + ({\bf v} - {\bf V})\nabla \varphi 
\nonumber\\
& = & ({\bf v} - {\bf V})(m \dot{\bf v} + \nabla \varphi)\nonumber\\ 
& = & - \frac{2}{3} \frac{e²}{c³} {\dot{\bf v}}².
\end{eqnarray}
Now the total force ${\bf f}_{tot}$ is the sum of the force ${\bf f}_{M\rightarrow m}$ 
which the source 
particle $M$ exerts on particle $m$, and the radiation reaction force ${\bf f}_{rad}$, 
i.e. ${\bf f}_{tot} = {\bf f}_{M\rightarrow m} + {\bf f}_{rad}$.
Using the equation of motion $m\dot{\bf v}= {\bf f}_{tot}$ and 
the relation $\nabla \varphi  = - {\bf f}_{M\rightarrow m}$, one gets 
$m \dot{\bf v} + \nabla \varphi = {\bf f}_{tot} - {\bf f}_{M\rightarrow m} = {\bf f}_{rad}$,
and we are left with

\begin{equation}
({\bf v} - {\bf V}){\bf f}_{rad} = -  \frac{2}{3} \frac{e²}{c³} {\dot{\bf v}}².
\end{equation}
The auxiliary potential $\varphi$ has completely disappeared.
Thus the equation for the  radiation force becomes
\begin{equation}
 {\bf f}_{rad} = -  \frac{2}{3} \frac{e²}{c³}{\dot{\bf v}}²\frac{{\bf v} - {\bf V}}{({\bf v} - {\bf V})²}.
\end{equation} 
One could have added a term ${\bf f}_\perp$ which is orthogonal to ${\bf v} - {\bf V}$. 
But from
the vectors which are available one can only construct 
$({\bf v} - {\bf V}) \times (\dot{\bf v} - \dot{\bf V})$
which is an axial vectors and therefore forbidden by parity.

Clearly invariance under non relativistic boosts is now restored, and the reason for the failure in the 
usual derivation is also clear. Performing the same steps as before, but eliminating 
the charged particle $m$ instead of the source $M$, one finds that
there is not only a force on the particle under consideration but also
a force on the source particle. Using the unphysical concept of an external field ignores one of the 
fundamental laws of physics: Isaac Newton's third law {\em actio = reactio}. 

The message learned from the above considerations is:
{\bf It is not possible to determine the radiation reaction force without knowing the motion
of the source which causes the acceleration.}

But now a curiosity arises, in our opinion not a paradox, but nevertheless a rather strange consequence.  
Consider, for instance, a linear potential, i.e. a constant force. This can be approximately created 
by a far away source $M$. Imagine we could measure the radiation reaction force
in a lab. Then, by comparison with (8), we could determine the relative velocity between
particle and source, even without seeing 
the source! The situation
becomes even more confusing if the force is due to several sources which are in motion with
respect to each other. Plenty of conceptual problems to think about.

We will first give two simple applications for one dimensional problems and compare 
our results with those from the non relativistic LAD equation (4). For applications 
we take, of course, the limit $M/m \rightarrow \infty$ and can accordingly  
choose a convenient system in which $X=const.$  and $V=0$, such that 
$\varphi$ becomes a function of $x$ only. 

The equation of motion for one dimensional problems then reads
\begin{equation}
m \ddot{x} =  - \varphi ' (x) - m \tau_c \frac{\ddot{x}²}{\dot{x}},
\end{equation}
where $\tau_c$ denotes the characteristic time 

\begin{equation}
\tau_c = \frac{2}{3} \frac{e²}{c³ m}, 
\end{equation}
not to be confused with the proper
time $\tau$ in the following section.
Solving the quadratic equation (9) for $\ddot{x}$ gives
\begin{equation}
 \ddot{x} = - \frac{\dot{x}}{2\tau_c} 
\left(1 \pm \sqrt{1 - \frac{4\tau_c \varphi'}{m\dot{x}}}\right)
\rightarrow \left\{ {-\dot{x}/\tau_c \atop - \varphi'/m}\right\} 
\mbox{ for } \tau_c \rightarrow 0.
\end{equation}
Only the lower sign for the square root gives the correct limit
for $\tau_c \rightarrow 0$, therefore we will always choose this
sign in the following. One also realizes the possibility of a 
complex square root. Note that the differential equation is non
linear, we don't have the possibility of constructing real 
solutions by superposition. We will clarify the origin of this phenomenon 
when treating the example of a constant external force.

\begin{center}
{\Large \em Constant external force }
\end{center}

We consider a charge in a field with constant acceleration $-g$. The equation of motion is 
(this equation was also used by Hammond \cite{Hammond}, without, however, 
mentioning the classically forbidden regions)
\begin{equation}
m\dot{v} = -mg -m\tau_c \frac{\dot{v}²}{v}, 
\end{equation}
or
\begin{equation} 
\dot{v} = - \frac{v}{2\tau_c }(1 - \sqrt{1 - 4 g \tau_c /v}).
\end{equation}
This is a first order separable differential equation for $v(t)$. 
It can be explicitly integrated to obtain $t$ as function of $v$, but the result is not very 
enlightening. A qualitative discussion is more informative.

For negative $v$ the differential equation is perfectly well behaved. This corresponds to the
 case that we simply drop the charge in the (say) gravitational field. For positive
 $v$, however, i.e. if we throw the charge upwards, there is no longer 
a real solution if $0 < v < v_c$,  where we introduced the critical velocity
\begin{equation}
 v_c= 4g \tau_c.
\end{equation}
This behavior, though very strange at first sight,
can be well understood. If the charge moves upwards it gains potential energy, furthermore it
has to provide the energy for the radiation. These two energies have to be compensated by
a corresponding loss of kinetic energy. But near the turning point the velocity is so small that it is
no longer possible to decrease the kinetic energy sufficiently. There is no longer a solution 
for the energy requirement, one would need a negative kinetic
energy. The mathematics consequently answers with a complex solution!  
For negative velocities there is no problem. When the particle falls
downwards it looses potential energy which can be used to provide the energy for the radiation.

Clearly classical physics breaks down here. We are confronted with a typical
tunneling problem, where the particle has to tunnel through a classically forbidden
region. We also expect a certain probability for reflection instead of transmission.
For all practical situations one may safely ignore this (extremely small!)
forbidden region.  We 
did not attempt a quantum mechanical calculation of the tunneling. The problem is 
awkward, due to the lack of a Lagrangian or Hamiltonian formulation.

The problematic of the classically forbidden  region disappears in a perturbative 
treatment of first order in $\tau_c$. More precisely, because (12) can be written 
in the parameter free form 
$d\tilde{v}/d\tilde{t} = -1- (d\tilde{v}/d\tilde{t})²/\tilde{v}$, 
with $\tilde{t} =t/\tau_c,\: \tilde{v} = v/g\tau_c$, this means that $t \gg \tau_c$
with 
$t$ chosen such that the turning occurs near $t=0$. One finds the perturbative solution 
\begin{equation} 
\dot{v}(t)=-g(1-\tau_c/t). 
\end{equation}
This implies that $x(t)$ is finite everywhere, $v(t)$ has a logarithmic singularity, 
and $\dot{v}(t)$ a pole.
If one keeps away from the classically forbidden region one has an excellent
approximation to the exact solution.

Of course one can also check the energy balance 
$\Delta E_{pot} + \Delta E_{kin} + \Delta E_{rad}=0$ in first order of $\tau_c$ explicitly. 
The frequently asked question 
``Where does the energy come from if the particle permanently radiates?'' 
has a simple answer. The radiation has the effect that, at a given position,
 the particle has a smaller velocity than it would have without radiation, 
or, formulated the other way round, it reaches a certain velocity only at a lower altitude.

We compare with the LAD equation (4),
$m\dot{v} = - mg + m\tau_c \ddot{v}$, or  $a = -g + \tau_c \dot{a}$.
It has the solution $a= -g + a_i \exp(t/\tau_c)$, i.e. either a non radiating 
solution for $a_i=0$, or a run away solution for $a_i \neq 0$.

\begin{center}
{\Large \em Harmonic oscillator }
\end{center}
The equation of motion is
\begin{equation}
m \ddot{x}= - m\omega² x - m\tau_c \ddot{x}²/\dot{x}. 
\end{equation}
An ansatz of the form $x(t) = x_i\exp(\lambda t)$ leads to two 
complex solutions but these are useless for constructing a real solution
because the equation is non linear and we are not allowed to build a superposition.

Remembering the previous discussion we expect that there will be
 no real solution close to the extrema where the velocity is small. 
 Again the problem is less drastic in lowest order perturbation theory. One finds

\begin{equation}
x(t) = x_0 \cos\omega t\exp[-\tau_c (\omega ²t/2 - \omega \tan\omega t \ln |\sin \omega t|)].
\end{equation}
We compare with the LAD equation
$m \ddot{x} = - m \omega² x + m \tau_c \stackrel{...}{x}$. 
This is easily solved by the ansatz $x(t)= x_0 \exp(\lambda t)$ with $\lambda$ a
solution of the cubic equation $\lambda ² + \omega ² - \tau_c \lambda³ = 0$.
There is always one positive solution for $\lambda$ which represents the run away solution.
Furthermore there is a pair of complex conjugate solutions with negative real part
which lead to  damped oscillations. If $\tau_c$ is treated in lowest order one finds 
$\lambda= \pm i \omega - \tau_c \omega ²/2$. This term is also present in our perturbative 
solution (17), but the latter contains further terms which take into account 
the varying radiation during
the oscillation. The LAD equation, on the other hand, averages the whole process. 

If we ignore the subtleties near the classically forbidden regions around the extrema we can 
derive a nice general property. Consider two consecutive zeros, located at 
$t_1$ and $t_2 = t_1 + T/2$. While $x_1=x_2=0$, the velocities are different, 
$v_2 = - \exp(-\alpha/2) v_1$, say. But since (16) is a second order differential
equation the solution is fixed by the initial values $x_i,v_i$. Furthermore, 
if $x(t)$ is a solution of (16), then this is also the case for $const \cdot x(t)$. 
Therefore one can immediately conclude $x(t + T/2)= -\exp(\alpha/2)\, x(t)$, or, consequently,
\begin{equation}
 x(t+nT/2)= (-1)^n \exp(-n\alpha/2)\, x(t) \mbox{ for any integer } n.
\end{equation}
Knowledge within one half period is sufficient to know the whole solution.

\begin{center}
{\Large \em Circular motion}
\end{center}

Consider a charged particle which is forced to move in a circle of
fixed radius $r$, and no other external forces present. 
For the radial and azimuthal components of velocity and acceleration one has, 
using the simplification $\dot{r}=0$,
\begin{equation}
 v_r =  0, \quad a_r  =  - v_\varphi²/r, \quad a_\varphi  = \dot{v}_\varphi.
\end{equation}
Therefore $a² = v_\varphi^4/r² + \dot{v}_\varphi²$.
This leads to the following differential equation for $v_\varphi \equiv v$:

\begin{equation}
m\dot{v} = -m \tau_c(\frac{v³}{r²} + \frac{\dot{v}²}{v}),
\end{equation}
or

\begin{equation}
\dot{v} = -\frac{v}{2 \tau_c} (1 - \sqrt{1 - 4\tau_c²v^2/r²)}.
\end{equation}
The radicand is positive for any non relativistic motion and 
macroscopic radius. We don't look for exact solutions but are 
content with an expansion in $\tau_c$ to first order where we
have to solve $\dot{v} = - \tau_c v³/r²$, resulting in

\begin{equation}
v = v_i/\sqrt{1 + 2\tau_c v_i² (t-t_i)/r²}. 
\end{equation}
In lowest order of $\tau_c$ and for $v \ll c$ this agrees with 
the result of Hammond \cite{Hammondrev} 
as well as with the solution from the LAD equation. 
This is not surprising because in this example approximately 
$\dot{\bf v} \perp {\bf v}$, therefore the neglection of the 
integrated terms 
$\dot{\bf v} {\bf v} |_{t_1}^{t_2}$ is justified.

\section{Relativistic equation}\
The relativistic generalization of the Larmor formula (1) reads
\begin{equation}
\frac{dP^\mu_{rad}}{d\tau} = -K u^\mu,
\end{equation}
with 

\begin{equation}
K= -\frac{2}{3} \frac{e²}{c^3}\frac{du_\nu}{d\tau} \frac{du^\nu}{d\tau}
= m \tau_c \, [(\frac{(d{\bf v}/d\tau)²}{1-v²} + \frac{({\bf v} d{\bf v}/d\tau)²}{(1-v²)²}],
\end{equation}
where $d\tau= \sqrt{1-v²}\, dt$ denotes the proper time of the particle, and 
$u^\mu = (1/\sqrt{1-v²})(1,{\bf v})$ the four-velocity.
With the exception of the coefficient in $K$ we put $c=1$ everywhere. 

We closely follow the non relativistic treatment. For deriving the formula 
we use a simple model with an instantaneous potential $\varphi ({\bf x} - {\bf X})$.
This model does not claim any physical relevance. It is translation invariant but  
not Lorentz invariant. Nevertheless it will lead us to a covariant formula for the 
reaction force. The potential $\varphi ({\bf x} - {\bf X})$ is only an auxiliary construct.
As in the non relativistic case it will completely disappear in the final formula.

The momentum and energy of our system are

\begin{eqnarray}
{\bf P} & = & \frac{m}{\sqrt{1-v²}}{\bf v} + \frac{M}{\sqrt{1-V²}}{\bf V},\\
P⁰ & = & \frac{m}{\sqrt{1-v²}} + \frac{M}{\sqrt{1-V²}} + \varphi ({\bf x} - {\bf X}).
\end{eqnarray}
Now apply the time derivative $d/dt = \sqrt{1-v²} \, d/d\tau$ to these equations. 
For the derivation we can specialize to a one dimensional motion where the 
formulae simplify. 
The differentiation can be easily performed, the result has to be identical to
$-K \sqrt{1-v²}\, u^\mu = -K(1,v)$. This leads to the two equations 

\begin{eqnarray}
\frac{m}{(1-v²)^{3/2}}\frac{dv}{dt} + \frac{M}{(1-V²)^{3/2}} \frac{dV}{dt} 
& = & - Kv,\\
\frac{m}{(1-v²)^{3/2}} v\frac{dv}{dt} + \frac{M}{(1-V²)^{3/2}} V\frac{dV}{dt}
+ (v-V) \varphi' 
& = & - K.
\end{eqnarray}
Eliminating the term with $M$ from the momentum equation (27) and introducing into the 
energy equation (28) one obtains

\begin{equation} 
(v-V) \Big(\frac{m}{(1-v²)^{3/2}} \frac{dv}{dt} + \varphi'\Big)  = -K(1-Vv).
\end{equation}
We now work with the four-forces. As in the non relativistic case we have 
$F_{tot} = F_{M \rightarrow m} + F_{rad}$, 
and we can use the equation of motion for $F_{tot}$ and the connection
between $\varphi'$ and $F_{M \rightarrow m}$,

\begin{equation} 
F_{tot} = \frac{m}{(1-v²)²} \frac{dv}{dt}, \mbox{ and } \varphi' = -\sqrt{1-v²} F_{M \rightarrow m},
\end{equation}
therefore
\begin{equation}
 \Big(\frac{m}{(1-v²)^{3/2}} \frac{dv}{dt} + \varphi'\Big) 
= \sqrt{1-v²} (F_{tot}- F_{M \rightarrow m}) = \sqrt{1-v²} F_{rad}.
\end{equation}
One is left with 

\begin{equation} 
F_{rad} = - K \frac{1 -Vv}{\sqrt{1-v²}(v-V)}.
\end{equation}
This can be written in a manifestly covariant form. 
In the non relativistic limit the force is 
proportional to the velocity, therefore $F^\mu_{rad}$ has to contain a term
proportional to the four-velocity $u^\mu$. In order to fulfill the condition 
$u_\mu F^\mu_{rad} =0$ it must appear in  the combination $u^\mu - U^\mu/Uu$.
 One has
\begin{equation}
u^m - U^m/Uu = \frac{v^m - V^m + v^2 V^m- {\bf Vv} v^m}{\sqrt{1-v^2}(1-{\bf V v})}
\end{equation}
and
\begin{equation}
(u^\mu - U^\mu/Uu)^2 = 
- \frac{({\bf v} - {\bf V})^2 -V^2v^2+({\bf Vv})^2}{(1 - {\bf V v})^2}.
\end{equation}
Therefore (32), which refers to the special case ${\bf V} \parallel {\bf v}$, 
can be written as the spatial part of
\begin{equation} 
F^\mu _{rad} =   \frac{K}{(u^\nu - U^\nu/uU)^2}(u^\mu - U^\mu/Uu),
\end{equation}
with $K$ given in (24).
It is important to emphasize that the formulation of $F^\mu _{rad}$ as a four-force
is only possible if one has the four-vector $U^\mu$ available. We are confident that
this solution is unique.

For applications we will of course again consider the limit 
$M/m \rightarrow \infty$, so we can put $U^\mu \approx (1,0)$. The 
spatial component of the reaction force (35) then simply becomes

\begin{equation} 
F^m_{rad} =  K \frac{u^m}{(u^\nu - U^\nu/Uu)²} = -\frac{K}{\sqrt{1-v²}}\frac{v^m}{v²}.
\end{equation}

We give a simple application.

\begin{center}
{\Large \em  Constant external electric field}
\end{center}

We consider a linear motion in a constant electric field $-E$
with $E>0$. The minus sign was chosen for convenient comparison with
the non relativistic case discussed previously. For 
${\bf v} \parallel \dot{\bf v}$ the formula (24) for $K$ simplifies to
$K = m\tau_c (dv/d\tau)²/(1-v²)²$, and
the equation of motion, conveniently written in terms of $t$,
not of $\tau$, reads

\begin{equation}
 m \frac{dv}{dt} = -(1-v²)^{3/2} eE - m \tau_c \frac{(dv/dt)²}{(1-v²)^{3/2} \, v}.
\end{equation}
This corresponds to the previous non relativistic equation (12) with the replacements
$g \rightarrow  (1-v²)^{3/2}eE/m, \: \tau_c \rightarrow \tau_c (1-v²)^{-3/2}$.
We can again solve for $dv/dt$ and, performing these replacements in (14), obtain 
the critical velocity $v_c =  4eE\tau_c/m$, and the classically forbidden
region $0<v < v_c$. 

The solution of (37) without the radiation term is well known. In first order 
perturbation theory in $\tau_c$ the solution of the full equation is

\begin{equation}
 v(t)=  -  \Big[\frac{eEt}{q}
 -  \tau_cmeE\big(\frac{1}{q²} +\frac{m}{2q^3}
\ln \frac{q-m}{q+m}\big)\Big],\nonumber
\end{equation}
with
\begin{equation}
 q = \sqrt{m²+(eE)²t²}.
\end{equation}
The limit for large times is particularly simple:

\begin{equation}
 v(t)\rightarrow -\big[ 1 - \frac{m²}{2(eE)²} \big(1 + \frac{2 \tau_ceE}{m}\big)\frac{1}{t²}\big].
\end{equation}
It shows how the approach of $v$ to $-1$ is slowed down by the radiation.
\newpage

\begin{center}
{\Large \em Summary and conclusions}
\end{center}

We summarize our essential results and the properties of the equations
derived here. 
\begin{itemize}
\item It is mandatory to take into account the presence
of the source which causes the acceleration of the charged particle. Only
then one ends up with a force with the correct transformation property.
\item There can be (extremely small) regions where the velocity 
is so small that the energy requirements cannot be fulfilled. 
The charge has to pass through such regions by quantum mechanical tunneling.
It is impressive how the equation responds
to the appearance of classically forbidden regions. 
\item Because we chose a direct approach without manipulations like partial 
integrations our equation does not contain third
 derivatives $\stackrel{...}{x}$,  in clear contrast to the LAD equation.
 Consequently there are no problems
with run away solutions. 
\item The non relativistic equation for the radiation reaction force can
be considered as unique, because it was derived in a direct 
straight forward way. We believe that this is also the case for the 
relativistic generalization.
\item In all examples considered so far we obtained 
reasonable results. 
\end{itemize}

\end{document}